\documentclass[prb,aps,preprint,showpacs,amssymb]{revtex4}

\usepackage{graphicx}

\begin{document}
  
  \title{Electron-phonon coupling in boron-doped diamond superconductor}

  \author{H. J. Xiang}

  \author{Zhenyu Li}
   
  \author{Jinlong Yang }

  \thanks{Corresponding author. E-mail: jlyang@ustc.edu.cn}

  \author{J. G. Hou}

  \author{Qingshi Zhu}

  \affiliation{Hefei National Laboratory for Physical Sciences at
    Microscale, Laboratory of Bond Selective Chemistry and Structure
    Research Laboratory, University of Science and Technology of
    China, Hefei, Anhui 230026, People's Republic of China}

  \date{\today}

  \begin{abstract}
    The electronic structure, lattice dynamics, and electron-phonon coupling
    of the boron-doped 
    diamond are investigated using the density functional supercell
    method.
    Our results
    indicate the boron-doped diamond is a phonon mediated
    superconductor,
    confirming previous theoretical conclusions deduced from the
    calculations employing the virtual crystal
    approximation. We show that the optical phonon modes involving B
    vibrations play an important role in the electron-phonon coupling.
    Different from previous 
    theoretical results, our calculated
    electron-phonon coupling
    constant is 0.39 and the estimated superconducting transition
    temperature $T_{c}$ is 4.4 K for 
    the boron doped diamond with $2.78\%$ boron content using the
    Coulomb pseudopotential $\mu^* = 0.10$,
    in excellent agreement with the experimental result. 
  \end{abstract}

  \maketitle

  Recently, Ekimov {\it et al.} \cite{B-C1}
  reported the discovery of superconductivity in the boron-doped diamond
  synthesized at high pressure and
  high temperature. Their measurements
  showed 
  that the boron-doped diamond with a hole carrier density of
  $5\times 10 ^{21}$ cm$^{-3}$
  is a bulk, type-II superconductor below the
  superconducting transition temperature $T_{c} \approx 4$ K. Using
  a simplified McMillan formula \cite{McMillan} by setting the
  Coulomb pseudopotential $\mu^* $ to zero, they estimated the       
  electron-phonon coupling (EPC) constant $\lambda=0.2$, indicative of weak
  EPC. 

  Before that, superconductivity in  
  some doped semiconductors has been discovered
  experimentally,\cite{SrTiO3,TlBiTe2} and 
  superconductivity in many-valley degenerate
  semiconductor has been also predicted theoretically. \cite{valley} 
  As the first group-IV diamond type semiconductor with 
  superconducting properties, the boron-doped diamond superconductor has been 
  studied by several groups.
  Baskaran {\it et al.} \cite{Baskaran} suggested that when the boron
  doping concentration increases to a critical value $n_c \approx 4
  \times 10^{21}/$cm$^3$, an Anderson-Mott insulator to resonating
  valence bond (RVB) \cite{RVB} superconductivity transition
  takes place.
  By first
  principles density functional perturbation calculations \cite{DFTP}
  employing the virtual crystal approximation (VCA),
  \cite{VCA} Boeri {\it et al.} \cite{Boeri} substantiated  that  the  
  recently discovered  
  superconductivity below 4 K in 3\% boron-doped diamond is caused by
  the coupling of a few
  holes at top of the $\sigma$-bonding valence band to the optical
  zone-center phonons, similar as in
  MgB$_{2}$, \cite{MgB2,MgB2_1,MgB2_2,MgB2_3} albeit 
  in 3 dimensions. 
  Another first principles study employing
  the VCA and frozen phonon method
  also indicated that the EPC 
  is the likely superconducting mechanism. \cite{Lee}
  Regardless of
  their similar conclusion in these two studies,
  we note that there are some discrepancies
  in these two studies, especially in the EPC
  strength and $T_c$: $\lambda=0.27$ and 
  $T_{c}=0.2$  K when $\mu^* = 0.10$ 
  for $3\%$ doping concentration in the first study,
  however, $\lambda=0.55$ and $T_{c}= 9 $ K when $\mu^* = 0.15$
  for $2.5\%$ doping concentration in the second study.
  The difference in doping concentration makes the discrepancy even more
  prominent since larger doping concentration leads to larger 
  $\lambda$ and $T_{c}$.
  Moreover, though qualitative agreement with experiment is found in these
  two studies, some obvious disagreements still exist, in particular,
  $T_{c}$ in the first study is too small, and $T_{c}$ in the second
  study is too large though one can obtain $T_{c}=4$ K by setting
  $\mu^*$ to a somewhat large value, i.e., 0.20.  
  The disagreements between experiment and theory might indicate the
  inapplicability of the VCA to
  the newly found boron-doped diamond superconductor.
  In fact there are some cases where the VCA failed,
  e.g., the VCA was inappropriate for
  calculating the band gaps of GaPN and GaAsN, \cite{GaPN} 
  and it described badly the structural and electronic properties of the
  quaternary alloy GaAlAsN. \cite{GaAlAsN} On the other hand, though
  computational demanding,
  the supercell method is usually reliable. \cite{GaPN,GaAlAsN}   
  To bridge the gap between experiment and theory, and to verify the
  phonon mediated superconducting mechanism in the boron-doped diamond,   
  here we report a first principles supercell calculation on the
  EPC of 
  the boron-doped diamond. Our results
  support the conventional phonon mediated superconducting mechanism
  in boron-doped diamond and the calculated $T_{c}$ is in excellent
  agreement with the experimental value.

  Electronic structure calculations and geometrical optimizations are 
  performed using density functional theory (DFT) \cite{DFT1,DFT2} in
  the local density approximation (LDA). \cite{LDA1,LDA2}
  The electron-ion interaction is described by ultrasoft
  pseudopotentials, \cite{uspp} which allow a low
  cutoff energy 
  (25 Ry in this work) in the plane-wave expansion.
  The phonon and EPC 
  calculations are carried out using density functional
  perturbation theory in the linear response. \cite{DFTP}   
  Within the phonon mediated theory of superconductivity,
  $T_{c}$ can be estimated 
  using McMillan's solution of the Eliashberg equation, \cite{McMillan}
  \begin{equation}
    T_{c}=\frac{\omega _{\ln  }^{ph}}{1.20}\exp \left\{ -\frac{1.04
      \left(1+\lambda \right)}{%
      \lambda -\mu ^{\ast }(1+0.62\lambda )}\right\} ,
    \label{mcm}
  \end{equation}
  where $\lambda $ is the EPC constant,
  $\omega _{\ln  }^{ph}$ is
  the logarithmically averaged characteristic phonon frequency,
  and 
  $\mu  ^{\ast }$  is the Coulomb pseudopotential which describes the
  effective electron-electron repulsion.    
  The EPC constant $\lambda$ is
  calculated as an average over the $N$ ${\bf q}-$points
  mesh and over all the phonon modes,  
  $\lambda=\sum_{{\bf q}\nu} \lambda_{{\bf q}\nu}/N$, where
  $\lambda_{{\bf q}\nu}$ is 
  the electron-phonon interaction for a phonon mode $\nu$ with
  momentum {\bf q}.   
  The modes responsible for superconductivity can be identified
  from the Eliashberg function, 
  \begin{equation}
    \alpha^2F(\omega)=\frac{1}{2 N}\sum_{{\bf q}\nu} \lambda_{{\bf
	q}\nu} \omega_{{\bf q}\nu} \delta(\omega-\omega_{{\bf q}\nu} ) .
  \end{equation}
  Then $\omega _{\ln  }^{ph}$ is
  calculated as, 
  \begin{equation}
    \omega _{\ln  }^{ph}=\exp \left\{ \frac{2}{\lambda }\int_{0}^{\infty }d\omega
    \alpha ^{2}F(\omega )\ln \omega /\omega\right\}.
  \end{equation}

\begin{figure}[tbp]
    \includegraphics[width=7cm]{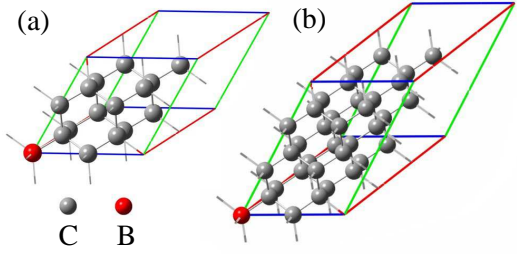}
    \caption{(Color online) Structures of (a) model I and (b) model
    II. Refer to the text 
    for the description for the models.}
    \label{fig1}
  \end{figure}

  We use the supercell technique to model the boron-doped diamond. 
  In order to study the dependence of the EPC on
  the doping concentration, we choose two models: 
  $2 \times 2  \times 2$ and  
  $3 \times 3 \times 2$ diamond supercells with a carbon atom
  substituted by a boron atom, named models I and II respectively.
  The total B content ($C_{B}$) for
  experimental samples is $2.8 \pm 0.5 \% $, which is smaller than the total B
  content ($6.25\%$) in model I and very close to the total B
  content ($2.78\%$) in model II.
  The two models are shown in Fig.~\ref{fig1}. The optimized lattice
  constant for diamond is 3.57 \AA, agreed well 
  with the experimental lattice constant (3.566 \AA) \cite{B-C1} and previous
  theoretical results. \cite{d_ph}
  For the boron-doped diamond
  we check that our calculations reproduce the slight lattice
  expansion, less than 0.3\%. \cite{B-C1} 
  Since the not very large boron doping has negligible
  effect on the lattice constant of diamond,
  we use our optimized lattice constant (3.57 \AA) in all subsequent
  calculations. 
  
  The k-point integration for geometrical optimization, construction of the
  induced charge density, and calculation of the dynamical matrix is
  performed over a  
  $4 \times 4 \times 4$ ($2 \times 2 \times 3$) Monkhorst-Pack grid
  \cite{mp} for  model I (II),
  and a finer  
  $8 \times 8 \times 8$  ($4 \times 4 \times 6$)  grid is used in the
  phonon linewidth calculations where the convergence in
  the k-point sampling is more
  difficult than that for the phonon calculations.
  The dynamical matrix and phonon linewidth are computed on a
  $3 \times 3 \times 3$  ($2 \times 2 \times 3$) {\bf q}-point mesh
  for model I (II), and a Fourier interpolation is used to
  obtain complete phonon dispersions.  
  
\begin{figure}
    \includegraphics[width=6cm]{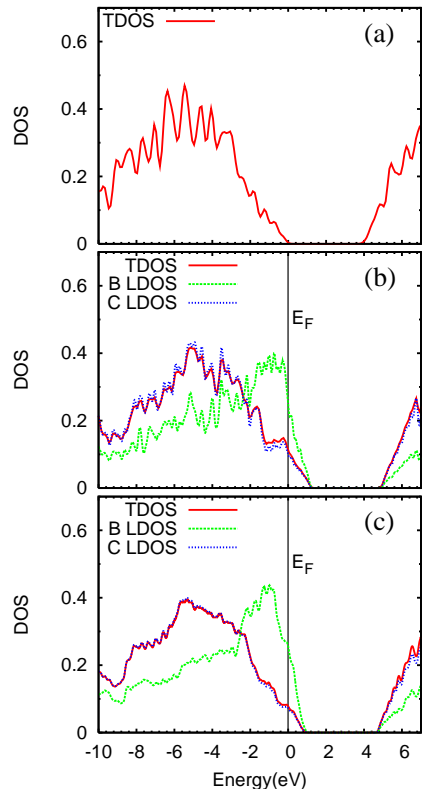}
    \caption{(Color online) Electronic DOS for the undoped diamond and
      two boron-doped diamond models. 
      (a) shows the TDOS of the undoped diamond, (b) and (c) are the
      TDOS and LDOS plots for models I and II.
      The LDOS for C shown here represents the average LDOS of all the C
      atoms.
      TDOS is in states/(spin $\cdot$ eV $\cdot$  diamond unit cell).
      The unit of LDOS is states/(eV $\cdot$ supercell). Energy
      is relative to the Fermi level $E_{F}$ (for the undoped diamond, 
      the valence top is taken as zero enegy point). } 
    \label{fig2}
  \end{figure}

  The L\"owdin population analysis has been carried out to get the local
  density of states (LDOS). 
  The total electronic density of states (TDOS) and LDOS for models I
  and II are plotted in  Fig.~\ref{fig2}  
  (The TDOS for the undoped diamond is also shown as a reference).
  The TDOS clearly indicates a degenerate or metallic behavior in
  both models, contrasting sharply to the semiconductor behavior of
  the undoped diamond.   
  The TDOS and LDOS for model I are very similar with those for model
  II except that the width of the acceptor bands (the bands between
  the valence top and the fermi level) for model
  I is larger than that for model II.
  For both cases, the width of the acceptor bands is larger than that
  for a boron-doped diamond 64-atom supercell,
  \cite{B-C2} suggesting a dependence of the
  width of the acceptor bands on the boron doping concentration. When
  the doping concentration 
  decreases, the width of the acceptor bands decreases. At very small
  doping concentration, the acceptor bands even no longer
  overlap with the valence band edge of the diamond resulting in a
  threefold degenerate acceptor state with a hole bind energy of
  $E_{B}\approx 0.37 $ eV. \cite{low_B} 
  Since there
  are only s and p electrons in boron doped diamond, the electron
  correlation should not be very strong. The metallization in
  the boron-doped diamond is reasonablely caused by the increased
  boron content, doesn't necessarily resort to
  the Anderson impurity model, \cite{s-d,B-C_metal} which is suited for
  describing strong correlated systems.
  Fig.~\ref{fig2}(b) and (c) contain the LDOS plots of the B atom
  and the average LDOS of the C atoms. 
  From Fig.~\ref{fig2} we can clearly see that the LDOS around
  the Fermi level for B is larger than that the average LDOS for the C
  atoms.   
  However, it 
  doesn't mean the electronic states near the Fermi level are localized
  around B since the sum of the LDOS for all C atoms is larger than
  the LDOS for B. 

\begin{figure}
    \includegraphics[width=8.5cm]{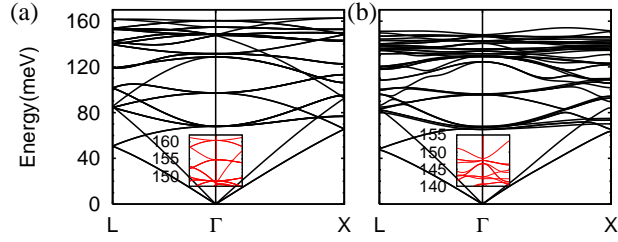}
    \caption{(Color online) Phonon band structures for (a) $2\times 2 \times 2$
    supercell diamond and (b) $2\times 2 \times 2$
    supercell diamond with a carbon atom substituted by a boron atom
    (model I). The insets show the high energy optical phonon branches
    near $\Gamma$ both for the undoped and doped systems.  }
    \label{fig3}    
  \end{figure}

  The calculated frequency of the highest optical phonon at
  $\Gamma$ for the updoped 
  diamond is 1295 cm$^{-1}$, agreed well with previous LDA
  calculations, \cite{Boeri,d_ph} a little smaller than experimental
  frequency, 1332 cm$^{-1}$. \cite{B-C1}
  The phonon band structures for the undoped and boron-doped 
  $2 \times 2 \times 2$ diamond supercell are shown in
  Fig.~\ref{fig3}(a) and (b) respectively. 
  Here, we plot the phonon band structure for the
  diamond supercell just for comparison with that for the boron-doped
  diamond. The phonon band structure
  for the diamond supercell
  is more complex than that for the unit cell
  since there are much more carbon atoms in the
  supercell.
  There are more branches for the boron-doped diamond since some phonon
  degenerates are removed due to the symmetry breaking, especially the
  highest optical phonon at $\Gamma$ is a one-fold A$_{1}$ mode, as shown
  in the inset of Fig.~\ref{fig3}(b).
  Beside that, a general effect to the phonon band structures
  resulting from  the boron doping is the phonon softening.  
  Especially the softening of the optical phonons
  is sizeable, e.g., the highest frequency of the  zone center
  optical phonons decreases  99 cm$^{-1}$ to 1196 cm$^{-1}$ for model I. 
  Comparing our results with those obtained from the VCA, we find that
  the softening of the highest optical
  phonon modes at $\Gamma$ in our supercell calculations is much
  smaller (the decrease in theirs is 265 cm$^{-1}$ for diamond with
  5\% boron content \cite{Boeri}).  
  Zhang {\it et al.} observed that the zone-center optical phonon
  line at 1332 cm$^{-1}$ downshifted for boron doped diamond film. \cite{Zhang}
  Part of the reason for the downshifting might be the softening of
  the optical phonon due to the EPC.
  We can see that there is a slight upturn of the uppermost mode
  especially when moving from $\Gamma$ to X for both updoped and
  doped diamond, similar with the previous results.
  \cite{phn_diam,d_ph} 
  Moreover, the upturn for the boron-doped diamond is more noticeable
  due to the larger EPC for the zone
  center optical phonons.       

\begin{figure}
    \includegraphics[width=8.5cm]{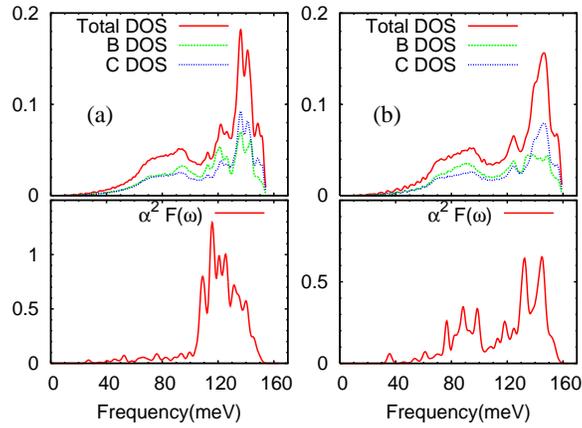}
    \caption{(Color online) Total phonon DOS, boron
      partial phonon DOS and average partial phonon DOS for all carbon
      atoms,
      and Eliashberg function 
      $\alpha^{2}F(\omega)$ 
      for (a) model I and (b) model II. 
      The unit for the total phonon DOS is states/(meV $\cdot$ diamond
      unit cell). 
      Partial phonon DOS shown here is in arb. unit.}
    \label{fig4}
  \end{figure}

  The partial phonon density of states (DOS) for atom $a$ is defined as: 
  $\rho_{a}(\omega)=\sum_{\bf q} \sum_{j=1}^{3N}|e_{a}({\bf
    q},j)|^{2}\delta(\omega-\omega({\bf q},j))$,
  where $N$ is the total number of atoms, ${\bf q}$ is the
  phonon momentum, $j$ labels the phonon branch, $e_{a}({\bf q},j)$ is
  the phonon  
  displacement vector for atom $a$, and $\omega({\bf q},j)$ is the phonon
  frequency.      
  The total phonon DOS, B partial phonon DOS
  and average partial phonon DOS of all C atoms
  for both models are shown in the 
  upper panels of Fig.~\ref{fig4}. 
  The shapes of the phonon DOS for
  these two models are very similar,
  while the phonon softening
  effect in model I is stronger due to the larger boron concentration. 
  The lower panels of
  Fig.~\ref{fig4} are the Eliashberg function  $\alpha^{2}F(\omega)$
  for both models. 
  The extremely weak signal in the low frequency part
  (lower than 80 meV) of $\alpha^{2}F(\omega)$ indicates very
  weak electron acoustic phonon coupling.   
  Detailed analysis shows that the B vibrations related EPC
  is the largest among all atoms in both models I and II (14.5\%
  in model I and 9.3\% in model II). 
  Thus the B related phonon modes play an important
  role in the EPC due to the large B electronic LDOS near the Fermi
  level.
  As shown in the lower panels of Fig.~\ref{fig4},
  $\alpha^{2}F(\omega)$ has sizeable contributions from 
  phonons with medium frequency in our results. However, Boeri {\it et
  al.} suggested that
  $\alpha^{2}F(\omega)$ vanishes for phonon
  frequencies below that of the optical zone-center modes, then jumps
  to a maximum, and finally falls. \cite{Boeri} Their
  relative simple picture for  
  $\alpha^{2}F(\omega)$ should stem from the fact that there are only
  three optical phonon modes in their virtual crystal calculations. 
  Since we involve much more phonon modes to calculate
  $\alpha^{2}F(\omega)$, our results should be more reliable.

  The calculated TDOS at the Fermi level ($N(E_{F})$), average frequencies
  $\omega _{\ln  }^{ph}$, EPC constants
  $\lambda$, and 
  superconducting  transition temperatures $T_{c}$ for models I and
  II are listed in Table~\ref{table1}. 
  As expected, the
  $N(E_{F})$ for model I is larger than that for model II  due to the
  larger boron concentration.   
  The $N(E_{F})$ for the boron-doped diamond is consistent with the
  virtual crystal calculations: $N(E_{F})$ (in states/(spin $\cdot$
  eV $\cdot$  diamond unit cell) is 0.062 for 2.78\% boron
  vs 0.060 for 2.5\% boron, \cite{Lee} 0.106 for 6.25\%
  boron vs 0.08 for 5\% boron. \cite{Boeri} The smaller $N(E_{F})$ for
  the boron-doped diamond  than  MgB$_{2}$ would result in weaker
  EPC in  the boron-doped diamond.   
  The large $\omega _{\ln  }^{ph}$ for
  the boron-dope diamond leads to high $T_{c}$, as demonstrated in
  McMillan's formula. \cite{McMillan}
  Thought the phonon softening increases with
  increased boron concentration, $\omega _{\ln  }^{ph}$ for model I
  is larger than that for model II, i.e., 1287 K {\it vs} 1218
  K. The smaller 
  $\omega _{\ln  }^{ph}$ for model II  results from the substantial
  EPC in the medium frequency region. 
  The EPC
  constant $\lambda$ for model I is 0.56, larger than that
  ($\lambda=0.39$) for model II.
  The typical value for $\mu
  ^{\ast }$  is in the range 0.10 to 0.15. The calculated $T_{c}$  for
  model I is 
  23.6 (11.5) K  for $\mu  ^{\ast }=$0.10 (0.15), 
  and for model II it is
  4.4 (0.9) K for $\mu  ^{\ast }=$0.10 (0.15).
  The boron  concentration for model II is very close to the
  experimental value and the calculated  $T_{c}$  agrees well with
  experimental $T_{c}$, about 4 K. We can see $T_{c}$ increases
  with increased boron concentration due to the increased
  $\lambda$. 
  Detailed analysis reveals that the EPC is
  peaked at $\Gamma$,
  which is also found in previous study. \cite{Boeri}
  Lee {\it et al.} \cite{Lee} overestimated $\lambda$ and $T_{c}$
  since $\lambda$  should be
  averaged over the whole BZ, but their frozen phonon calculations
  only took into account the zone center phonons. 
  
  Our treatment neglects the anharmonic corrections.
  For MgB$_{2}$, Choi {\it et al.} \cite{MgB2_2} showed that $\lambda$
  is decreased from 1.0 to 0.78 after including partial anharmonic
  corrections. However, more recently, Lazzeri {\it et al.}
  \cite{MgB2_3} suggested that the effects 
  of anharmonicity on $T_{c}$ and $\lambda$ in MgB$_{2}$ are indeed
  negligible by explicitly taking into account the scattering between
  different phonon modes at different {\bf q} points in the whole BZ.
  Boeri {\it et al.} \cite{Boeri}  
  considered the anharmonic corrections in the boron-doped diamond using
  frozen phonon calculations and found that the effect of anharmonicity is
  small, i.e., in general $\lambda$ decreases 0.03 after taking
  into account the anharmonicity. So our main results
  remain essentially unchanged even after including the anharmonic
  corrections since the anharmonic corrections for the boron-doped
  diamond are small, as discussed by Lee {\it et al.}. \cite{Lee} 
  
  In summary, we have carried out a first principles study on the
  heavily boron-doped diamond.  
  Optical phonons in
  diamond are softened after doping with boron. 
  The boron related vibrational modes contribute an important part to
  the Eliashberg function $\alpha^{2}F(\omega)$.
  Superconductivity in the boron-doped diamond is
  found to be mediated by the EPC. 
  By using the supercell technique, we resolve the
  discrepancy between theoretical results based on the VCA
  and experimental data, and the calculated
  $T_{c}$ is in excellent agreement with the experimental result. 
  
  This work is partially supported by the National Project for the
  Development of Key Fundamental Sciences in China (G1999075305,
  G2001CB3095), by the National Natural Science Foundation of China
  (50121202, 20025309, 10074058), 
  and by the USTC-HP HPC project.

  \clearpage

  \begin{table}
    \caption{Computed electronic DOS at the Fermi level ($N(E_{F})$),
      average frequencies 
      $\omega _{\ln  }^{ph}$, EPC
      constants $\lambda$, and 
      superconducting  transition temperatures $T_{c}$ for models I and
      II. The two
      values for $T_{c}$ correspond to two different values of 
      $\mu  ^{\ast}$ (0.10 and 0.15). $N(E_{F})$ is in
      states/(spin $\cdot$ eV $\cdot$  diamond unit cell). Boron content
      $C_{B}$ is also 
      shown for each model.} 
    \begin{tabular}{cccccc}
      \hline
      \hline
      & $C_{B}$  &$N(E_{F})$&$\omega _{\ln  }^{ph}$ (K)&$\lambda$&
      $T_{c}$ (K)\\ 
      \hline
      Model I& 6.25\%  & 0.106 &1287 & 0.56 & 23.6,11.5\\
      Model II& 2.78\% & 0.062 &1218 &0.39 & 4.4,0.9\\
      \hline
      \hline
    \end{tabular}
    \label{table1}
  \end{table}

  \clearpage

\end{document}